# Thermohaline circulation stability: a box model study - Part I: uncoupled model

Valerio Lucarini *

and

Peter H. Stone †

Department of Earth, Atmospheric and Planetary Sciences, MIT

Cambridge, MA 02138 USA

and

Joint Program on the Science and Policy of Global Change, MIT

Cambridge, MA 02138 USA

September 25, 2004

---

*lucarini@alum.mit.edu; now at Department of Mathematics and Computer Science, University of Camerino, via Madonna delle Carceri 62032, Camerino (MC), Italy

†phstone@mit.edu

**Abstract**

A thorough analysis of the stability of the uncoupled Rooth inter-hemispheric 3-box model of thermohaline circulation (THC) is presented. The model consists of a northern high latitudes box, a tropical box, and a southern high latitudes box, which respectively correspond to the northern, tropical and southern Atlantic ocean. We adopt restoring boundary conditions for the temperature variables and flux boundary conditions for the salinity variables. We study how the strength of THC changes when the system undergoes forcings that are analogous to those of global warming conditions by applying to the equilibrium state perturbations to the moisture and heat fluxes into the three boxes. In each class of experiments, we determine, using suitably defined metrics, the boundary dividing the set of forcing scenarios that lead the system to equilibria characterized by a THC pattern similar to the present one, from those that drive the system to equilibria with a reversed THC. Fast increases in the moisture flux into the northern high-latitude box are more effective than slow increases in leading the THC to a breakdown, while the increases of moisture flux into the southern high-latitude box strongly inhibits the breakdown and can prevent it, in the case of slow increases in the Northern Hemisphere. High rates of heat flux increase in the North Hemisphere destabilize the system more effectively than low ones, and increases in the heat fluxes in the Southern Hemisphere tend to stabilize the system.

# 1. Introduction

Following the classical picture of Sandström's, the ocean circulation can be divided, albeit somewhat ambiguously, into the wind-driven circulation, which is directly induced by the mechanical action of the wind and essentially affects the surface waters, and the thermohaline (THC) circulation, which is characterized by relatively robust gradients in the buoyancy of the water masses (Weaver and Hughes 1992; Rahmstorf 2000, 2002; Stocker et al. 2001). Although some studies show that the THC is very sensitive to wind stresses (Toggweiler and Samuels 1995), more recent studies suggest that the buoyancy forces are dominant when more realistic boundary conditions are considered (Rahmstorf and England 1997; Bugnion and Hill 2003). The THC dominates the large scale picture of the global ocean circulation and is especially relevant for the intermediate-to-deep water.

One of the main issues in the study of climate change is the fate of the THC in the context of global warming (Weaver and Hughes 1992; Manabe and Stouffer 1993; Stocker and Schmittner 1997; Rahmstorf 1997, 1999b,a, 2000; Wang et al. 1999a,b). The THC plays a major role in the global circulation of the oceans as pictured by the conveyor belt scheme (Weaver and Hughes 1992; Stocker 2001); therefore the effect on the climate of a change of its pattern may be global (Broecker 1997; Manabe and Stouffer 1999b; Cubasch et al. 2001; Seager et al. 2002). The THC is sensitive to changes in the climate since the North Atlantic deep water (NADW) formation is sensitive to variations in air temperature and in precipitation in the Atlantic basin (Rahmstorf and Willebrand 1995; Rahmstorf 1996).

There are several paleoclimatic data sets indicating that changes in the patterns or collapses of the THC coincide with large variations in climate, especially in the North Atlantic region (Broecker et al. 1985; Boyle and Keigwin 1987; Keigwin et al. 19994; Rahmstorf 2002).

Box models have historically played a major role in the understanding of the fundamental dynamics of the THC (Weaver and Hughes 1992). The Stommel two-box oceanic model (Stommel 1961) parameterized the THC strength as proportional to the density gradient between the northern latitude and equatorial boxes. The oceanic model proposed by Rooth (1982) introduced the idea that the driver of the THC was density difference between the high latitude basins of the northern and southern hemispheres, thus implying that the THC is an inter-hemispheric phenomenon. Both the Stommel and Rooth models allow two stable equilibria, one - the warm mode - characterized by downwelling of water in the high latitudes of the North Atlantic, which resembles the present oceanic circulation, and the other - the cold mode - characterized by upwelling of water in the high latitudes of the North Atlantic. Perturbations to the driving parameters of the system, *i.e.* freshwater and heat fluxes in the oceanic boxes, can cause transitions between the two regimes.

Various GCM experiments have shown that multiple equilibria of the THC are possible (Bryan 1986; Manabe and Stouffer 1988; Stouffer et al. 1991; Stocker and Wright 1991; Manabe and Stouffer 1999a; Marotzke and Willebrand 1991; Hughes and Weaver 1994), and most GCMs have shown that the GHGs radiative forcing could cause a weakening of the THC by the inhibition of the sinking of the water in the Northern Atlantic. Large increases of the moisture flux and/or of the surface air temperature in the downwelling regions are the driving mechanisms of such a process. Nevertheless in a recent global warming simulation run the THC remained almost constant since the destabilizing mechanisms have been offset by the increase of the salinity advected into the downwelling region due to net freshwater export from the whole North Atlantic (Latif et al. 2000).

Hemispheric box models along the lines of Stommel's have been extensively studied and the stability of their THC has been assessed in the context of various levels of complexity in the representation of the coupling between the ocean and the atmosphere (Weaver and Hughes 1992; Nakamura et al. 1994; Marotzke 1996; Krasovskiy and Stone 1998; Tziperman and Gildor 2002).

However, analysis of GCM data have indicated that the THC is an interhemispheric phenomenon, and in some cases it has been found that the THC strength is approximatively proportional to the density difference between the Northern and Southern Atlantic (Hughes and Weaver 1994; Rahmstorf 1996; Klinger and Marotzke 1999; Wang et al. 1999a), thus supporting Rooth's approach.

In this paper we perform an analysis of the stability of the present pattern of the THC against changes to the heat and freshwater atmospheric fluxes using an uncoupled version of the Rooth oceanic model. We note that this model assumes that buoyancy is well mixed in the vertical, and thus avoids the need for parameterizing the mixing (Munk and Wunsch 1998). The model here considered has no real explicit coupling between the ocean and the atmosphere: the atmospheric freshwater fluxes are prescribed and the atmospheric heat fluxes relax the oceanic temperatures towards prescribed values (Rahmstorf 1996; Scott et al. 1999; Stone and Krasovskiy 1999; Titz et al. 2002b,a).

We explicitly analyze what is the role of the spatial patterns *and* of the rates of increase of the forcings in determining the response of the system. We consider a suitably defined metric and determine which are the thresholds beyond which we have destabilization of the warm mode of the THC. We underline that our treatment goes beyond quasi-static analysis since the effect of the rate of forcing is explicitly addressed, so that we analyze how the effects on the system of slow (Wang et al. 1999a) and rapid (Wiebe and Weaver 1999) changes join on; in particular in the limit of very slow changes, our results coincide with those that could be obtained with the analysis of the bifurcations of the system (Stone and Krasovskiy 1999; Scott et al. 1999). Relatively few studies with more complex models have explicitly addressed how the spatial (Rahmstorf 1995, 1996, 1997) or temporal (Stocker and Schmittner 1997) patterns of forcing determine the response of the system. However, these studies did not carry out a parametric study of the influence of both spatial *and* temporal patterns. We emphasize that our study is based on a process model which aims at obtaining a qualitative picture of the properties of the THC more than providing quantitative results.

Our paper is organized as follows. In section 2 we provide a description and the general mathematical formulation of the dynamics of the three-box model used in this study. In section 3 we describe the parameterization of the physical processes. In section 4 we analyze the feedbacks of the system. In section 5 we describe the stability properties of the system when it is forced with perturbations in the freshwater flux. In section 6 we describe the stability properties of the system when it is forced with perturbations in the heat flux. In section 7 we present our conclusions.

## 2. Model description

The three-box model consists of a northern high latitude box (box 1), a tropical box (box 2), and a southern high latitude box (box 3). The volume of the two high latitudes boxes is the same, and is $1/V$ times the volume of the tropical box. We choose $V = 2$, so that box 1, box 2 and box 3 respectively can be thought as describing the portions of an ocean like the Atlantic north of $30°N$, between $30°N$ and $30°S$, and south of $30°S$. At the eastern and western boundaries of the oceanic boxes there is land; our oceanic system spans $60°$ in longitude, so that it covers $\epsilon = 1/6$ of the total surface of the planet. The boxes are $5000\ m$ deep, so that the total mass $M_{i=1,3} = M$ of the water contained in each of the high-latitude boxes is $\approx 1.1 \cdot 10^{20}\ Kg$, while $M_2 = V \cdot M_{i=1,3} = 2 \cdot M$. The tropical box is connected to both high latitude boxes; moreover the two high latitude boxes are connected by a deep current passage (which bypasses the tropical box) containing a negligible mass. The three boxes are assumed to be well-mixed. The assumption of well-mixed boxes excludes the *polar halocline disaster* (Bryan 1986; Weaver and Hughes 1992; Zhang et al. 1993) from the range of phenomena that can be described by this model. In our model we have not included a deep box as done in Rahmstorf (1996). This choice allows an easier and more detailed control of the behavior of the system, but implies that distortions are introduced in the response of the system to fast perturbations (Weaver and Hughes 1992). The physical state of the box $i$ is described by its temperature $T_i$ and its salinity $S_i$; the box $i$ receives from the atmosphere the net flux of heat $\tilde{H}_i$ and the net flux of freshwater $\tilde{F}_i$; the freshwater fluxes $\tilde{F}_i$ globally sum up to $0$, since the freshwater is a globally conserved quantity. The box $i$ is subjected to the oceanic advection of heat and salt from the upstream box through the THC, whose strength is $\tilde{q}$.

In figure 1 we present a scheme of our system in the northern sinking pattern: note that the arrows representative of the freshwater fluxes are arranged in such a way that the conservation law is automatically included in the graph. The dynamics of the system is described by the tendency equations for the heat and the salinity of each box. We divide the three heat tendency equations by $c_p \cdot M_i$, where $c_p$ is the constant pressure specific heat of water per unit of mass, and in the salinity tendency equations we neglect the contribution of the freshwater fluxes in the mass balance, so that virtual salinity fluxes and freshwater fluxes are equivalent (Marotzke 1996; Rahmstorf 1996; Scott et al. 1999). We obtain the following final form for the temperature and salinity tendency equations for the three boxes (Scott et al. 1999):

$$\dot{T}_1 = \begin{cases} q(T_2 - T_1) + H_1, & q \geq 0 \\ q(T_1 - T_3) + H_1, & q < 0 \end{cases} \qquad (1)$$

$$\dot{T}_2 = \begin{cases} \frac{q}{V}(T_3 - T_2) + H_2, & q \geq 0 \\ \frac{q}{V}(T_2 - T_1) + H_2, & q < 0 \end{cases} \quad (2)$$

$$\dot{T}_3 = \begin{cases} q(T_1 - T_3) + H_3, & q \geq 0 \\ q(T_3 - T_2) + H_3, & q < 0 \end{cases} \quad (3)$$

$$\dot{S}_1 = \begin{cases} q(S_2 - S_1) - F_1, & q \geq 0 \\ q(S_1 - S_3) - F_1, & q < 0 \end{cases} \quad (4)$$

$$\dot{S}_2 = \begin{cases} \frac{q}{V}(S_3 - S_2) - F_2, & q \geq 0 \\ \frac{q}{V}(S_2 - S_1) - F_2, & q < 0 \end{cases} \quad (5)$$

$$\dot{S}_3 = \begin{cases} q(S_1 - S_3) - F_3, & q \geq 0 \\ q(S_3 - S_2) - F_3, & q < 0 \end{cases} \quad (6)$$

where $q = \rho_0 \cdot \tilde{q}/M$, $H_i = \tilde{H}_i/(c_p \cdot M)$, and $F_i = \rho_0 \cdot S_0 \cdot \tilde{F}_i/M$.

We impose that the average salinity is a conserved quantity; this means that $\dot{S}_1 + V\dot{S}_2 + \dot{S}_3 = 0$, which implies that:

$$F_2 = -\frac{1}{V}(F_1 + F_3). \quad (7)$$

This conservation law holds at every time, and so rules out the possibility of including in our study the effects of the melting of continental ice sheets. We note that the system can (asymptotically) reach an equilibrium only if its feedbacks can drive it towards a state in which the following equation holds:

$$H_2 = -\frac{1}{V}(H_1 + H_3). \tag{8}$$

The strength of the specific THC is parameterized as proportional to the difference between the density of the water contained in the box $1$ and the density of the water contained in the box $3$, as in the Rooth (1982) model. Given that the water density can approximately be expressed as:

$$\rho(T, S) \approx \rho_0(1 - \alpha T + \beta S); \tag{9}$$

where $\alpha$ and $\beta$ are respectively the thermal and haline expansion coefficients, set respectively to $1.5 \cdot 10^{-4}\ °C^{-1}$ and $8 \cdot 10^{-4}\ psu^{-1}$, and $\rho_0$ is a standard reference density, we obtain for the normalized THC strength $q$ the following relation:

$$q = \frac{k}{\rho_0}(\rho_1 - \rho_3) = k\left[\alpha\left(T_3 - T_1\right) + \beta\left(S_1 - S_3\right)\right], \tag{10}$$

where $k$ is the hydraulic constant, chosen to obtain a reasonable value of the THC strength. We emphasize that the constant $k$ roughly summarizes the wind-powered turbulent mixing taking place in the ocean and powering the THC, and we think that in principle it should be expressed as a function of the climatic state of the system rather than being a constant.

The northern (southern) sinking state of the circulation is therefore characterized by $q > (<)0$. Considering the case $q > 0$, we obtain a diagnostic relation for steady state $q$ by setting to 0 equations (3) and (6), and using equation (9):

$$q^{eq} = [k\,(\alpha H_3^{eq} + \beta F_3^{eq})]^{1/2}\,; \qquad (11)$$

in the case $q < 0$, a diagnostic relation for the steady state $q$ can be obtained using the same procedure as above but setting to zero equations (1) and (4):

$$q^{eq} = -\,[k\,(\alpha H_1^{eq} + \beta F_1^{eq})]^{1/2}\,. \qquad (12)$$

We conclude that the *equilibrium* value of the THC strength is determined by the equilibrium values of the heat and freshwater fluxes of the box where we have upwelling of water. These results generalize the expressions given in (Rahmstorf 1996). The transient evolution of $q$ is determined by its tendency equation:

$$\begin{aligned}\dot{q} &= -q^2 + kq\,[\alpha\,(T_1 - T_2) - \beta\,(S_1 - S_2)] + \\ &\quad + k\,[\alpha\,(H_3 - H_1) + \beta\,(F_3 - F_1)] \qquad,\quad q \geq 0,\end{aligned} \qquad (13)$$

$$\begin{aligned}\dot{q} &= q^2 + kq\,[\alpha\,(T_3 - T_2) - \beta\,(S_3 - S_2)] + \\ &\quad + k\,[\alpha\,(H_1 - H_3) + \beta\,(F_1 - F_3)] \qquad,\quad q < 0,\end{aligned} \qquad (14)$$

We observe that the difference between the forcings applied to the freshwater and heat fluxes into the two boxes 1 and 3 directly effect the evolution of $q$; the presence of terms involving the gradient of temperature and salinity between box 2 and box 1 (box 3) if $q > 0$ ($q < 0$) breaks the symmetry of the role played by the two high latitude boxes. In our experiments we perturb an initial equilibrium state by changing at various rates the parameters controlling the fluxes $H_i$ and/or $F_i$ and observe under which conditions we obtain a reversal of the THC. The reversal of the THC causes a sudden cooling and freshening in the box 1 and a sudden warming and increase of the salinity in the box 3, because the former loses and the latter receives the advection of the tropical warm and salty water.

## 3. Parameterization of the physical processes

We perform our study considering the original formulation of the Rooth (1982) model and thus consider *mixed* boundary conditions for the surface fluxes. For a more detailed description of the model see also the paper by Scott et al. (1999). The heat flux into the box $i$ is described by a newtonian relaxation law of the form $H_i = \lambda_i(\tau_i - T_i)$ where the parameter $\tau_i$ is the target temperature (Bretherton 1982), which represents a climatologically reasonable value towards which the box temperature is relaxed, and the parameter $\lambda_i$ quantifies the efficacy of the relaxation. Along the lines of the paper of Scott et al. (1999) and following the parameterization introduced by Marotzke (1996), we choose $\tau_1 = \tau_3 = 0°C$ and $\tau_2 = 30° C$, and select $\lambda_1 = \lambda_2 = \lambda_3 = \lambda = 1.29 \cdot 10^{-9}\ s^{-1}$, which mimics the combined effect of radiative heat transfer and of a meridional heat transport pa-rameterized as being linearly proportional to the meridional temperature gradient (Marotzke and Stone 1995; Marotzke 1996; Scott et al. 1999). In physical terms, $\lambda$ corresponds to a restoring coefficient of $\approx 4.3\ Wm^{-2}\ °C^{-1}$ affecting the whole planetary surface, which translates into an effective $\tilde{\lambda} \approx 1/\epsilon \cdot 4.3\ W\ m^{-2}\ °C^{-1} = 25.8\ W\ m^{-2}\ °C^{-1}$ relative to the oceanic surface fraction only.

Using equations (1-3), we can obtain the following tendency equation for the globally averaged temperature $T_M = (T_1 + VT_2 + T_3)/(2+V)$:

$$\dot{T}_M = \lambda(\tau_M - T_M) \qquad (15)$$

where $\tau_M$ is the average target temperature; the average temperature relaxes towards its target value in about $25\ y$, which is very close to the value considered by Titz et al. (2002b); we observe that in the case of an ocean with vertical resolution, our heat flux parameterization corresponds to a restoring time of $\approx 3$ months for a 50 m deep surface level (Marotzke and Stone 1995).

The virtual salinity fluxes $F_i$ are not functions of any of the state variables of the system, and are given parameters. We choose $F_1 = 13.5 \cdot 10^{-11}\ psu\ s^{-1}$ and $F_3 = 9 \cdot 10^{-11}\ psu\ s^{-1}$, which respectively correspond to net atmospheric freshwater fluxes of $\approx 0.40\ Sv$ into the box 1 and of $\approx 0.27\ Sv$ into the box 3. These values are based on the analysis of the hydrology of the two hemispheres presented by Baumgartner and Reichel (1975) and by Peixoto and Oort (1992). Our model neglects moisture fluxes associated with the wind-driven circulation. If these were strong enough to reverse the sign of $F_3$, the model's behavior would be very different (Rahmstorf 1996; Titz et al. 2002b), but would not be in accord with observations.

Choosing $k = 1.5 \cdot 10^{-6} \ s^{-1}$, at equilibrium we have $q \approx 1.47 \cdot 10^{-10} \ s^{-1}$ (Scott et al. 1999). In physical terms this value describes a THC in the northern sinking pattern with a strength of $\approx 15.5 \ Sv$: this value agrees reasonably well with estimates coming from observations (Roemmich and Wunsch 1985; Macdonald and Wunsch 1996). We emphasize that this THC strength introduces an oceanic time scale $t_s \approx 250y$, which is the ratio between the box 1 and 3 volumes and the equilibrium state water transport. We refer to this time scale as the *flushing* or *advective time* of the ocean since it is the time it takes for the ocean circulation to empty out box 1 or box 3 as well as to propagate the information from one box to another. We underline that, since we have considered the approximation of well-mixed boxes, our model should not be considered very reliable for phenomena taking place in time scales $\ll t_s$.

In this equilibrium state boxes 1 and 3 receive a net poleward oceanic heat advection of $\approx 1.58 \ PW$, and $\approx 0.17 \ PW$ respectively. The former agrees reasonably well with estimates. The latter agrees with the sign of the global transport in the Southern Hemisphere, but disagrees with the sign in just the Southern Atlantic (Peixoto and Oort 1992; Macdonald and Wunsch 1996). The equilibrium temperature of box 1 is larger than the equilibrium temperature of box 3, and the same inequality holds for the salinities, the main reason being that box 1 receives directly the warm and salty tropical water. Taking into account equation (9) and considering that the actual equilibrium $q$ is positive, we conclude that the THC is haline-dominated: the ratio between the absolute value of the thermal and of the haline contribution on the right side of equation 9 is $\approx 0.8$. The occurrence of the haline dominated mode depends of course on the forcing parameters, which we have based on observations. The haline mode, with its positive value for q, is consistent with observations. We report in table 1 the values of the main model constants, while in table 2 we present the fundamental parameters characterizing this equilibrium state.

## 4. Feedbacks of the system

The newtonian relaxation law for the temperature implies that the atmosphere has a negative thermic feedback:

- $T_i$ increases $\Rightarrow$ $H_i$ decreases $\Rightarrow$ $T_i$ decreases

The reaction of the ocean to a decrease in the strength of the THC can be described as follows:

- $q$ decreases $\Rightarrow$

1. $T_1$ decreases more than $T_3$ $\Rightarrow$
    (a) $q$ increases
    (b) the change is limited by the atmospheric feedback;

2. $S_1$ decreases more than $S_3$ $\Rightarrow$ $q$ decreases;

The heat advection feedback (1) is negative and dominates the positive salinity advection feedback (2). We underline that the sign of the salinity feedback depends on the asymmetry in the freshwater fluxes Scott et al. (1999). We note that the strength of the second order feedback mechanism ($1b$) is relevant in establishing the overall stability of the THC (Tziperman et al. 1994); a very strong temperature-restoring atmospheric feedback like that in our model tends to decrease the stability of the system (Nakamura et al. 1994; Rahmstorf 2000). The feedbacks $(1) + (2)$ are governed by $q$, so that their time scale is around the flushing time $t_s \approx 250\ y$ (Scott et al. 1999) of the oceanic boxes. We expect that a fast perturbation avoids those feedbacks, since it bypasses the ocean-controlled transfer of information needed to activate them.

In order to simulate global warming conditions, we perform two sets of experiments. In the first, we increase the freshwater fluxes into the two high-latitude boxes. The rationale for this forcing is in the fact that an increase in the global mean temperature produces an increase in the moisture capacity of the atmosphere, and this is likely to cause the enhancement of the hydrological cycle, and especially in the transport of moisture from the tropics to the high latitudes. In the second, we represent the purely thermic effects of global warming by setting the increase in the target temperatures of the two high-latitude boxes larger than that of the tropical boxes, since virtually all GCM simulations forecast a larger increase in temperatures in the high latitudes (the so-called *polar amplification*).

# 5. Freshwater flux forcing

We force the previously defined equilibrium state with a net freshwater flux into the two high latitudes boxes which increases linearly with time; this is obtained by prescribing:

$$F_i(t) = \left( \begin{array}{ll} F_i(0) + F_i^t \cdot t, & 0 \leq t \leq t_0 \quad i = 1, 3 \\ F_i(0) + F_i^t \cdot t_0 & t > t_0 \quad\quad i = 1, 3 \end{array} \right. \quad (16)$$

We impose the conservation of the globally averaged salinity of the oceanic system by requiring that equation (7) hold at all times. When the perturbation is over, we reach a final state that can be either qualitatively similar to the initial northern sinking equilibrium or radically different, *i.e.* presenting a reversed THC. Each perturbation can be uniquely specified by a set of three parameters, such as the triplet 1 $[t_0, \Delta F_3/\Delta F_1, \Delta F_1]$, or the triplet 2 $[t_0, \Delta F_3/\Delta F_1, F_1^t]$, where $\Delta F_i \equiv F_i^t \cdot t_0$ is the total change of the freshwater flux into box $i$. In order to describe a global warming scenario, we explore the effect of positive $\Delta F_1$ and $\Delta F_3$. Freshwater forcing into box 1 tends to destabilize the THC, since it induces a freshening and so a decrease of the density of the water. If the freshwater forcing is larger in the box 3, we do not obtain a reversal of the THC, the main reason being that increases in the net freshwater flux into box 3 tend to stabilize the circulation, as can be seen in equation (11). Since we are interested in the destabilizing perturbations, we limit ourselves to the case $\Delta F_3/\Delta F_1 < 1$. These perturbations cause an initial decrease in the THC strength $q$ and so trigger the

following feedbacks:

- $F_1$'s increase larger than $F_3$'s $\Rightarrow$ $S_1$ decreases more than $S_3$ $\Rightarrow$ $q$ decreases.

We present in figure 2 two trajectories of $q$ describing the evolution of the system from the initial equilibrium state when two different forcings, one subcritical and one supercritical, are applied. The oscillations in the value of $q$, which are slowly decaying in the subcritical case, are caused by the action of the heat and salinity advection feedbacks, which act on time scales of the order of the flushing time of the oceanic boxes ($t_s \approx 250$ years). In the case of the supercritical perturbation the oscillations of the value of $q$ grow in size until the system collapses to a southern sinking equilibrium. The two trajectories separate $\approx 400$ years after the beginning of the perturbation, when the negative feedback due to the decrease of advection of warm, salty water in box 1 becomes of critical relevance. The threshold value for $q$ is of the order of $10\ Sv$: more complex models have shown that weak THC are not stable (Rahmstorf 1995; Tziperman 2000). When the THC does not change sign, it recovers to its initial unperturbed value (if $\Delta F_3 = 0$) or overshoots it ($\Delta F_3 > 0$), as can be deduced from 11; this reminds us of results obtained with much more complex models (Wiebe and Weaver 1999).

**a.** *Hysteresis*

The hysteresis describes the response of the THC against quasi-static freshwater flux perturbations, *i.e.*, with a very large $t_0$, in both boxes 1 and 3. In figure 3 we have plotted 4 curves, corresponding to $\Delta F_3/\Delta F_1 = [0, 0.1, 0.2, 0.3]$. The obtained graphs essentially represent the stationary states for given values of the freshwater flux. For a given curve, each point $(x, y)$ plotted represents a stable state of the system having $q = y$ and $F_1 = x$. We scale all the graphs so that the common initial equilibrium state is the point $(1, 1)$. For a range of $F_1$, which depends on the value of $\Delta F_3/\Delta F_1$, the system is bistable, *i.e.* it possesses two distinct stable states, one having $q > 0$, the other one having $q < 0$. The history of the system determines which of the two stable states is realized. The boundaries of the domain in $F_1$ where the system is bistable correspond to subcritical Hopf bifurcations (Scott et al. 1999), which drive the system from the northern (southern) sinking equilibrium to the southern (northern) sinking equilibrium if $F_1$ crosses the right (left) boundary of the domain of bistability. If $\Delta F_3/\Delta F_1$ is large enough slow increases of the freshwater fluxes do not destabilize the system, so that no bifurcations are present. It is interesting to note that if $F_3$ is kept constant, the value of $q$ doe not depend on the value of $F_1$ in the northern sinking branch, while clearly the dependence is apparent in the southern sinking branch. This can be explained by considering that at equilibrium the value of $q$ is mainly determined by the value of the freshwater flux into the upwelling box, as shown in equation (11). We remark that since these hysteresis experiments involve simultaneous perturbations to the moisture fluxes in both hemispheres, they differ from those presented in Rahmstorf (1996), where the perturbations to the moisture fluxes occurred in one high-latitude box.

Albeit with differences in the experimental settings in some cases, the presence of an hysteresis for slow freshwater flux perturbations has been previsouly presented both for simple (Rahmstorf 1996; Scott et al. 1999; Titz et al. 2002b) as well as more complex interhemispheric models (Stocker and Wright 1991; Mikolajewicz and Maier-Reimer 1994; Rahmstorf 1995, 1996), while the inclusion of changes in the freshwater flux into both high latitude boxes as presented here had not been previously considered.

**b.** *Critical Forcings*

Figures 4a) and 4b) present, using respectively the coordinates described by the triplets 1 and 2, the contours of the height of the manifold dividing the subcritical from the supercritical forcings. We consider logarithmic scales instead of linear scales since the former allow to capture more clearly the qualitative behavior of the system. In figure 4a) we observe that, as expected the lower is the value of the ratio $\Delta F_3/\Delta F_1$, the lower is the total change $\Delta F_1$ needed to obtain the reversal of the THC. For a given value of the ratio $\Delta F_3/\Delta F_1$, more rapidly increasing perturbations (larger $F_1^t$) are more effective in disrupting the circulation. For values of $\Delta F_3/\Delta F_1 \leq 0.4$ we see a changeover between a *slow* and a *fast* regime, geometrically described by a relatively high- gradient region dividing two portions of the manifold having little $x-$dependence. Figure 4b) shows more clearly that for large values of $\Delta F_3/\Delta F_1$ ($\geq 0.5$) the collapse of the THC occurs only for fast increases, because of the presence of a threshold in the rate of increase depicted by the little $t_0-$dependence of the manifold. The changeover corresponds to $t_0 \approx t_s$, which essentially matches the flushing time of the oceanic boxes (Scott et al. 1999). For low values of $\Delta F_3/\Delta F_1$ the disruption of the initial THC pattern takes place even in the case of a very low $F_1^t$, as essentially captured in the hysteresis study of the previous subsection. This occurs because, as shown in the analysis performed by Scott et al. (1999), there is a critical ratio $F_3/F_1$ below which the northern sinking equilibrium becomes unstable. Such a ratio can be reached after a long enough $t_0$ if the slow perturbation is characterized by a low enough value of $\Delta F_3/\Delta F_1$.

### c. *Relevance of the mixed boundary conditions*

In order to explore the relevance of the mixed BCs for the stability of the THC pattern against freshwater flux perturbations (Tziperman et al. 1994; Nakamura et al. 1994), we have considered also a second version of the model where we replace the expression of the heat flux in the box $i$ defined as $\lambda(\tau_i - T_i)$ with a free parameter $H_i$, thus realizing a flux BC model. Choosing the value of the initial $H_i$ equal to the value of $\lambda(\tau_i - T_i)$ of the previous version of the model at equilibrium, we obtain the same equilibrium solution. We then perturb this equilibrium solution with freshwater flux forcing as in the previous case. Since the heat flux $H_i$ does not depend on $T_i$, the positive feedback ($1b$) described at the beginning of the section is off, therefore we expect this version of the system to be more stable than the previous one. Repeating the analysis of stability to freshwater flux perturbations for the flux BC model, it is indeed confirmed that temperature restoring conditions decrease the stability of the model, in agreement with the results of Tziperman et al. (1994), Nakamura et al. (1994), and Rahmstorf (2000).

Comparing the hysteresis graphs obtained for the mixed and flux BCs systems presented in figure 5, it is apparent that the mixed BCs model is less stable with respect to freshwater flux perturbations. Excluding the temperature feedback mechanism, we do not obtain collapses of the THC for quasi-static perturbations for $\Delta F_3/\Delta F_1 = 0.3$, while for each of the cases $\Delta F_3/\Delta F_1 = [0, 0.1, 0.2]$ the bifurcation point is much farther than in the mixed BC model from the point $(1,1)$, which describes the unperturbed system.

Figure 6 shows the ratio between the values of the critical values of $\Delta F_1$ presented in figure 4 and the corresponding values obtained with the flux BC model. The average value of this ratio over all the domain is $\approx 0.3$. Observing figure 6 for $\Delta F_3/\Delta F_1 = [0.3, 0.4]$, we see that the ratio goes to $0$ with increasing $t_0$ because in this subdomain for slow forcings we have destabilization for finite values of $\Delta F_1$ only in the mixed BCs case. This occurs because for these values the threshold in the rate of increase is present only in the flux BC model, therefore in this model for large $t_0$ the quantity $\Delta F_1$ diverges, while in the mixed BC model it is finite.

## 6. Thermal forcing

We explore the stability of the northern sinking equilibrium against perturbations to the initial heat fluxes expressed as changes in the target temperatures in the three boxes. We do not impose a constraint fixing the average target temperature $\tau_M$, so that in this case not two but three parameters $\tau_1, \tau_2, \tau_3$ can be changed. It is possible to recast the system of equations (1-6) (Scott et al. 1999) so that the heat forcing is described completely by the changes in $\tau_N = \tau_1 - \tau_2$ and in $\tau_S = \tau_3 - \tau_2$. Therefore, we set $\tau_2 = \tau_2(0)$ at all times, so that $\tau_1^t = \tau_N^t$ and $\tau_3^t = \tau_S^t$. We apply the following forcings:

$$\tau_i(t) = \begin{pmatrix} \tau_i(0) + \tau_i^t \cdot t, & 0 \leq t \leq t_0 & i = 1,3 \\ \tau_i(0) + \tau_i^t \cdot t_0 & t > t_0 & i = 1,3 \end{pmatrix} \quad (17)$$

We characterize each forcing experiment, which leads to a final state that is characterized by a northern or a southern sinking equilibrium, using the parameter set 1 defined as $[t_0, \Delta\tau_3/\Delta\tau_1, \Delta\tau_1]$ and the parameter set 2 defined as $[t_0, \Delta\tau_3/\Delta\tau_1, \tau_1^t]$; consistently with the previous case we define $\Delta\tau_i \equiv \tau_i^t \cdot t_0$, that is the total change of the target temperature of the box $i$. In order to imitate global warming scenarios where polar amplification occurs, we consider $\Delta\tau_i \geq 0$, $i = 1, 3$. We observe that if $\Delta\tau_3 > \Delta\tau_1$, no destabilization of the THC can occur, because such a forcing reinforces the THC. The specified forcings trigger the previously described feedbacks of the system in the following way:

- $\tau_1$'s increase larger than $\tau_3$'s $\Rightarrow$ $H_1$'s increase larger than $H_3$'s $\Rightarrow$ $T_1$ increases more than $T_3$ $\Rightarrow$ q decreases.

We present in figure 7 two trajectories of $q$ starting from the initial equilibrium state, characterized by slightly different choices of the forcing applied to $\tau_1$ and $\tau_3$. One of the forcing is supercritical, the other one is subcritical. The two trajectories are barely distinguishable up to the end of the perturbation, because the forcing dominates the internal feedbacks of the system thanks to the large value of $\lambda$. The oscillations in the value of $q$ at the end of the perturbation are caused by the internal feedbacks of the system and have a period comparable with the time scale of the flushing of the oceanic boxes. In general, the mean flow dampens the increase in $T_1$ and tends to keep the system in the northern sinking state, although the decreasing $q$ lowers the amount of warm, salty tropical water injected into box 1. When $q$ gets too small (in this case about $4\ Sv$) the mean flow feedback becomes negligible and the system eventually collapses.

**a.** *Hysteresis*

We present in figure 8 only the graphs relative to $\Delta\tau_3/\Delta\tau_1 = [0, 0.2, 0.5, 0.8]$ for sake of simplicity; the other curves that can be obtained for other values of $0 \leq \Delta\tau_3/\Delta\tau_1 < 1$ are not qualitatively different. The initial state is the point $(0, 1)$. The subcritical Hopf bifurcation points are respectively ($\approx 5$, $\approx 0.7$), ($\approx 6$, $\approx 0.7$)($\approx 9$, $\approx 0.7$), and ($\approx 18, \approx 0.6$). This means that the bifurcation occurs when the THC has already declined down to $\approx 10\ Sv$. The bistable region is extremely large in all cases, because the $x$ of the bifurcation points in the lower part of the circuits are below $-50°C$; they become farther and farther from the initial equilibrium value as $\Delta\tau_3/\Delta\tau_1$ increases. This means that once the circulation has reversed, the newly established pattern is extremely stable and can hardly be changed again. To our knowledge, this is the first time such an analysis of hysteresis graphs for this class of forcing is presented in the literature.

**b.** *Critical Forcings*

Figure 9 presents the contours of the height of the manifold of the critical forcings. We can observe that for all the values of $\Delta\tau_3/\Delta\tau_1$ there is a high-gradient region dividing two flat regions describing the *fast* and *slow* regimes. In the fast region, a smaller total increase $\Delta\tau_1$ is required to achieve the reversal of the THC. For $\Delta\tau_3/\Delta\tau_1 = 0.9$, which represents the case of highly symmetrical forcings, the critical $\Delta\tau_1$ essentially does not depend on $t_0$. We underline that the time scale of such a changeover is larger than $t_s \approx 250\,y$, the main reason being that because of the large value of $\lambda$, which is about one order of magnitude larger than the initial $q$, the forcing is stronger than any negative feedback even if it is relatively slow. For the same reason, in the case of perturbations to the target temperatures, we do not observe the presence of thresholds for the rate of increase of the perturbations as seen for freshwater forcings. If the increase in $\tau_1$ is slow enough, the negative feedbacks make the destabilization more difficult, but do not prevent it.

# 7. Conclusions

In this paper we have accomplished a thorough analysis of the stability of the THC as described by the Rooth model. We simulate global warming conditions by applying to the equilibrium state perturbations to the moisture and heat fluxes into the three boxes. In particular the relevance of the role of changes in the freshwater flux into the Southern Ocean is determined, along the lines of Scott et al. (1999) for a similar box model and Wang et al. (1999a) for an OGCM.

We have presented the hysteresis graphs for freshwater flux perturbations that reproduce and extend the results previously given in the literature, because the effect of freshwater flux increases in the Southern ocean are included, while for the first time hysteresis experiments for heat flux forcings, obtained by perturbation in the target temperature, are presented.

High rates of increase in the moisture flux into the northern high-latitude box lead to a THC breakdown at smaller final increases than low rates, while the presence of moisture flux increases into the southern high-latitude box strongly inhibit the breakdown. Similarly, fast heat flux increases in the North Hemisphere destabilize the system more effectively than slow ones, and again the enhancement of the heat fluxes in the Southern Hemisphere tend to drive the system towards stability. In all cases analyzed, slow forcings, if asymmetric enough, lead to the reversal of the THC.

In our model the spatial pattern of forcing is critical in determining the response of the system: quasi-symmetric forcings, which seem more reasonable under realistic global warming conditions, are not very effective in destabilizing the THC. The simplicity of the feedbacks characterizing our model, and in particular the absence of stratification effects, may be responsible for this behavior.

We have shown that the restoring boundary conditions for temperature decrease the stability of the system, because they introduce an additional positive feedback that tends to destabilize the system. Flux boundary conditions provide the system with much greater stability with respect to perturbations in the freshwater fluxes at all time scales.

Our study is based on a conceptual model, therefore it aims at obtaining a qualitative picture of the properties of the THC more than providing realistic results. Nevertheless, future increases in the computer power availability could allow the adoption of this methodology for performing extensive parametric studies of the THC stability properties and evolution with more complex and realistic models in both paleoclimatic and global warming perspectives.

We conclude that the study of the THC response to perturbations performed with uncoupled models, of any level of complexity, should take into much greater account the spatial pattern of the freshwater and/or heat flux forcing. The THC is a highly nonlinear, nonsymmetric, system, and we suggest that the effect of changing the rate of increase of the forcings should be explored in greater detail, to provide a bridge between the instantaneous and quasi-static changes. The THC dynamics encompasses very different time-scales, which can be explored only if the timing of the forcings is varied.

We conclude by pointing out some possible improvements to the present model and possible extensions of this work. We have put emphasis on how the oceanic advection can reverse the symmetry of the pattern of the circulation in the context of a rigorously symmetric geometry. The system is likely to be very sensitive to changes in the volumes of the boxes and especially to the introduction of asymmetries between the two high-latitude boxes, which would also make the system more realistic since the southern mid-high latitude portion of the Atlantic is considerably larger than the northern mid-high latitude portion (Zickfeld et al. 2003). Asymmetries in the oceanic fractional areas would induce asymmetries in the values of $\lambda_i$, thus causing the presence of different restoring times for the various boxes.

The introduction of a noise component in the tendency equation would increase the model's realism and make the model conceptually more satisfying since it would introduce a parameterization of the high-frequency variability. Recent studies have undertaken this strategy and shown that close to instability threshold the evolution of the THC has a very limited predictability (Wang et al. 1999a; Knutti and Stocker 2001), and that stochastic resonance could be responsible for glacial/interglacial climate shifts (Ganopolski and Rahmstorf 2002). It would be interesting to analyze how the intensity and the color of the noise would influence the results obtained and the conclusions drawn in this work.

Finally, the model could be improved so that it could be able to represent the main features of the whole conveyor belt, by adding other boxes, descriptive of other oceanic basins, and carefully setting the connections between the boxes, along the lines of some examples presented by Weaver and Hughes (1992). Moreover, the consideration of a deep ocean box would surely increase the model performance in terms of representation of the short time scales.

In an extension of this paper (Lucarini and Stone 2004) we perform a similarly structured stability analysis of the THC under global warming conditions using a more complex version of the Rooth model, where an explicit coupling between the oceanic and the atmospheric part is introduced and the main atmospheric physical processes responsible for freshwater and heat fluxes are formulated separately.

## Acknowledgments

The authors are grateful to J. Scott for interesting discussions and useful suggestions. One author (V.L.) wishes to thank R. Stouffer for having proposed improvements to an earlier version of the manuscript, and T. Stocker and E. Tziperman for having suggested a number of relevant references. This research was funded in part by the US Department of Energy's (DOE) Climate Change and Prediction Program and in part by the MIT Joint Program on the Science and Policy of Global Change (JPSPGC). Financial support does not constitute an endorsement by DOE or JPSPGC of the views expressed in this article.

# References

Baumgartner, A. and E. Reichel, 1975: *The World Water Balance*. Elsevier, New York.

Boyle, E. A. and L. Keigwin, 1987: North Atlantic thermohaline circulation during the past 20000 years linked to high-latitude surface temperature. *Nature*, **330**, 35–40.

Bretherton, F. P., 1982: Ocean climate modeling. *Progr. Oceanogr.*, **11**, 93–129.

Broecker, W. S., 1997: Thermohaline circulation, the Achilles heel of our climate system: Will man-made CO2 upset the current balance? *Science*, **278**, 1582–1588.

Broecker, W. S., D. M. Peteet, and D. Rind, 1985: Does the ocean-atmosphere system have more than one stable mode of operation? *Nature*, **315**, 21–26.

Bryan, F., 1986: High-latitude salinity effects and interhemispheric thermohaline circulations. *Nature*, **323**, 301–304.

Bugnion, V. and C. Hill, 2003: The equilibration of an adjoint model on climatological time scales. Part I: Sensitivity of meridional overturning circulation to the surface boundary conditions. Part II: The sensitivity of the thermohaline circulation to surface forcing and mixing in coupled and uncoupled models. Technical Report 72, Center for Global Change Science - MIT.

Cubasch, U., G. A. Meehl, G. J. Boer, R. J. Stouffer, M. Dix, A. Noda, C. A. Senior, S. Raper, K. S. Yap, A. Abe-Ouchi, S. Brinkop, M. Claussen, M. Collins, J. Evans, I. Fischer-Bruns, G. Flato, J. C. Fyfe, A. Ganopolski, J. M. Gregory, Z.-Z. Hu, F. Joos, T. Knutson, R. Knutti, C. Landsea, L. Mearns, C. Milly, J. F. B. Mitchell, T. Nozawa, H. Paeth, J. Risnen, R. Sausen, S. Smith, T. Stocker, A.Timmermann, U. Ulbrich, A. Weaver, J. Wegner, P. Whetton, T. Wigley, M. Winton, , and F. Zwiers: 2001, Projections of future climate change. *Climate Change 2001: The Scientific Basis. Contribution of Working Group I to the Third Assessment Report of the Intergovernmental Panel on Climate Change*, Cambridge University Press, Cambridge, 526–582.

Ganopolski, A. and S. Rahmstorf, 2002: Abrupt glacial climate changes due to stochastic resonance. *Phys. Rev. Lett.*, **88**, 038051.

Hughes, T. C. M. and A. J. Weaver, 1994: Multiple equilibrium of an asymmetric two-basin model. *J. Phys. Ocean.*, **24**, 619–637.

Keigwin, L. D., W. B. Curry, S. J. Lehman, and S. J. S., 19994: The role of the deep ocean in North Atlantic climate change between 70 and 130 ky ago. *Nature*, **371**, 323–326.

Klinger, B. A. and J. Marotzke, 1999: Behaviour of double hemisphere thermohaline flow in a single basin. *J. Phys. Ocean.*, **29**, 382–399.

Knutti, R. and T. F. Stocker, 2001: Limited predictability of the future thermohaline circulation close to an instability threshold. *J. Climate*, **15**, 179–186.

Krasovskiy, Y. and P. H. Stone, 1998: Destabilization of the thermohaline circulation by atmospheric transports: An analytic solution. *J. Climate*, **11**, 1803–1811.

Latif, M., E. Roeckner, U. Mikolajewicz, and R. Voss, 2000: Tropical stabilization of the thermohaline circulation in a greenhouse warming simulation. *J. Climate*, **13**, 1809–1813.

Lucarini, V. and P. H. Stone, 2004: Thermohaline circulation stability: a box model study - Part II: coupled model. *J. Climate*, in press.

Macdonald, A. M. and C. Wunsch, 1996: An estimate of global ocean circulation and heat fluxes. *Nature*, **382**, 436–439.

Manabe, S. and R. J. Stouffer, 1988: Two stable equilibria coupled ocean-atmosphere model. *J. Climate*, **1**, 841–866.

— 1993: Century-scale effects of increased atmospheric $CO_2$ on the ocean-atmosphere system. *Nature*, **364**, 215–218.

— 1999a: Are two modes of thermohalince circulation stable? *Tellus*, **51A**, 400–411.

— 1999b: The role of thermohaline circulation in climate. *Tellus*, **51A-B(1)**, 91–109.

Marotzke, J.: 1996, Analysis of thermohaline feedbacks. *Decadal Climate Variability: Dynamics and predicatibility*, Springer, Berlin, 333–378.

Marotzke, J. and P. H. Stone, 1995: Atmospheric transports, the thermohaline circulation, and flux adjustments in a simple coupled model. *J. Phys. Ocean.*, **25**, 1350–1360.

Marotzke, J. and J. Willebrand, 1991: Multiple equilibria of the global thermohaline circulation. *J. Phys. Ocean.*, **21**, 1372–1385.

Mikolajewicz, U. and E. Maier-Reimer, 1994: Mixed BC's in ocean general circulation models and their influence on the stability of the model's conveyor belt. *J. Geophys. Res.*, **99**, 22633–22644.

Munk, W. and C. Wunsch, 1998: Abyssal recipes II. Energetics of the tides and wind. *Deep-Sea Res.*, **45**, 1976–2009.

Nakamura, M., P. H. Stone, and J. Marotzke, 1994: Destabilization of the thermohaline circulation by atmospheric eddy transports. *J. Climate*, **7**, 1870–1882.

Peixoto, A. and B. Oort, 1992: *Physics of Climate*. American Institute of Physics, Washington.

Rahmstorf, S., 1995: Bifurcations of the Atlantic thermohaline circulation in response to changes in the hydrological cycle. *Climatic Change*, **378**, 145–149.

— 1996: On the freshwater forcing and transport of the Atlantic thermohaline circulation. *Clim. Dyn.*, **12**, 799–811.

— 1997: Risk of sea-change in the atlantic. *Nature*, **388**, 825–826.

Rahmstorf, S.: 1999a, Rapid oscillation of the thermohaline ocean circulation. *Reconstructing ocean history: A window into the future*, Kluwer Academic, New York, 309–332.

Rahmstorf, S., 1999b: Shifting seas in the greenhouse? *Nature*, **399**, 523–524.

— 2000: The thermohaline ocean circulation - a system with dangerous thresholds? *Climatic Change*, **46**, 247–256.

— 2002: Ocean circulation and climate during the past 120,000 years. *Nature*, **419**, 207–214.

Rahmstorf, S. and M. H. England, 1997: Influence of southern hemisphere winds on north atlantic deep water flow. *J. Phys. Oceanogr.*, **27**, 2040–2054.

Rahmstorf, S. and J. Willebrand, 1995: The role of temperature feedback in stabilizing the thermohaline circulation. *J. Phys. Ocean.*, **25**, 787–805.

Roemmich, D. H. and C. Wunsch, 1985: Two transatlantic sections: Meridional circulation and heat flux in the subtropical North Atlantic ocean. *Deep Sea Res.*, **32**, 619–664.

Rooth, C., 1982: Hydrology and ocean circulation. *Progr. Oceanogr.*, **11**, 131–149.

Scott, J. R., J. Marotzke, and P. H. Stone, 1999: Interhemispheric thermohaline circulation in a coupled box model. *J.Phys.Oceanogr.*, **29**, 351–365.

Seager, R., D. S. Battistini, J. Yin, N. Gordon, N. Naik, A. C. Clement, and M. A. Cane, 2002: Is the Gulf Stream responsible for Europe's mild winters? *Q. J. R. Meteorol. Soc.*, **128**, 2563–2586.

Stocker, T. F.: 2001, The role of simple models in understanding climate change. *Continuum Mechanics and Applications in Geophysics and the Environment*, Springer, Heidelberg, 337–367.

Stocker, T. F., R. Knutti, and G.-K. Plattner: 2001, The future of the thermohaline circulation - a perspective. *The Oceans and Rapid Climate Change: Past, Present and Future*, AGU, Washington, 277–293.

Stocker, T. F. and A. Schmittner, 1997: Influence of $CO_2$ emission rates on the stability of the thermohaline circulation. *Nature*, **388**, 862–864.

Stocker, T. F. and D. G. Wright, 1991: Rapid transitions of the ocean's deep circulation induced by changes in the surface water fluxes. *Nature*, **351**, 729–732.

Stommel, H., 1961: Thermohaline convection with two stable regimes of flow. *Tellus*, **13**, 224–227.

Stone, P. H. and Y. P. Krasovskiy, 1999: Stability of the interhemispheric thermohaline circulation in a coupled box model. *Dyn. Atmos. Oceans*, **29**, 415–435.

Stouffer, R. J., S. Manabe, and K. Bryan: 1991, Climatic response to a gradual increase of atmospheric carbon dioxide. *Greenhouse-Gas-Induced Climatic*

*Change: A Critical Appraisal of Simulations and Observations*, Elsevier, The Netherlands, 545–557.

Titz, S., T. Kuhlbrodt, and U. Feudel, 2002a: Homoclinic bifurcation in an ocean circulation box model. *Int. J. Bif. Chaos*, **12**, 869–875.

Titz, S., T. Kuhlbrodt, S. Rahmstorf, and U. Feudel, 2002b: On freshwater-dependent bifurcations in box models of the interhemispheric thermohaline circulation. *Tellus A*, **54**, 89–98.

Toggweiler, J. R. and B. Samuels, 1995: Effect of drake passage on the global thermohaline circulation. *Deep-Sea Res. I*, **42**, 477–500.

Tziperman, E., 2000: Proximity of the present-day thermohaline circulation to an instability threshold. *J. Phys. Ocean.*, **30**, 869–875.

Tziperman, E. and H. Gildor, 2002: The stabilization of the thermohaline circulation by the temperature/precipitation feedback. *J. Phys. Ocean.*, **32**, 2707–2714.

Tziperman, E., R. J. Toggweiler, Y. Feliks, and K. Bryan, 1994: Instability of the thermohaline circulation with respect to mixed boundary conditions: Is it really a problem for realistic models? *J. Phys. Ocean.*, **24**, 217–232.

Wang, X., P. H. Stone, and J. Marotzke, 1999a: Thermohaline circulation. Part I: Sensitivity to atmospheric moisture transport. *J. Climate*, **12**, 71–82.

— 1999b: Thermohaline circulation. Part II: Sensitivity with interactive atmospheric transport. *J. Climate*, **12**, 83–92.

Weaver, A. J. and T. M. C. Hughes: 1992, Stability of the thermohaline circulation and its links to climate. *Trends in Physical Oceanography*, Council of Scientific Research Integration, Trivandrum, 15.

Wiebe, E. C. and A. J. Weaver, 1999: On the sensitivity of global warming experiments to the parametrisation of sub-grid scale ocean mixing. *Clim. Dyn.*, **15**, 875–893.

Zhang, S., R. J. Greatbatch, and C. A. Lin, 1993: A re-examination of the polar halocline catastrophe and implications for coupled ocean-atmosphere modeling. *J. Phys. Ocean.*, **23**, 287.

Zickfeld, K., T. Slawig, and S. Rahmstorf, 2003: A low order model for the response of the Atlantic thermohaline circulation to climate change. *Ocean Dyn.*, In press.

# List of Tables

# List of Figures

| Quantity | Symbol | Value |
|---|---|---|
| Mass of Box $i = 1, 3$ | M | $1.08 \cdot 10^{20}\ Kg$ |
| Box2/Box$i = 1, 3$ mass ratio | V | 2 |
| Average water density | $\rho_0$ | $1000\ Kg\ m^{-3}$ |
| Specific heat per unit mass of water | $c_p$ | $4 \cdot 10^3\ J\ °C^{-1} Kg^{-1}$ |
| Average Salinity | $S_0$ | $35\ psu$ |
| Oceanic fractional area | $\epsilon$ | 1/6 |
| Thermal expansion coefficient | $\alpha$ | $1.5 \cdot 10^{-4}\ °C^{-1}$ |
| Haline expansion coefficient | $\beta$ | $8 \cdot 10^{-4}\ psu^{-1}$ |
| Hydraulic constant | k | $1.5 \cdot 10^{-6}\ s^{-1}$ |
| Atmospheric Freshwater Flux - Box 1 | $\tilde{F}_1$ | 0.41 Sv |
| Atmospheric Freshwater Flux - Box 2 | $\tilde{F}_2$ | -0.68 Sv |
| Atmospheric Freshwater Flux - Box 3 | $\tilde{F}_3$ | 0.27 Sv |
| Target Temperature - Box 1 | $\tau_1$ | $0\ °C$ |
| Target Temperature - Box 2 | $\tau_2$ | $30\ °C$ |
| Target Temperature - Box 3 | $\tau_3$ | $0\ °C$ |
| Atmospheric temperature restoring coefficient[a] | $\tilde{\lambda}$ | $25.8\ W m^{-2}\ °C^{-1}$ |

Table 1: Value of the main model constants

[a] Value relative to the oceanic surface fraction only

| Variable | Box 1 | Box 2 | Box 3 |
|---|---|---|---|
| Temperature | 2.9 $°C$ | 28.4 $°C$ | 0.3 $°C$ |
| Salinity | 34.7 psu | 35.6 psu | 34.1 psu |
| Surface Heat Flux | -1.58 PW | 1.74 PW | -0.16 PW |
| Oceanic Heat Flux | 1.58 PW | -1.74 PW | 0.16 PW |
| THC strength | 15.5 Sv | 15.5 Sv | 15.5 Sv |

Table 2: Value of the fundamental variables of the system at the initial equilibrium state

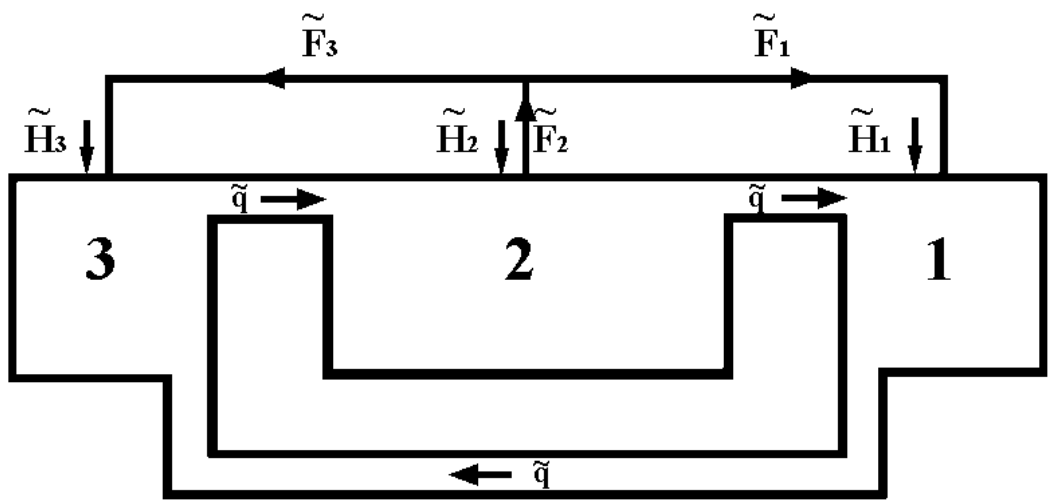

Figure 1: Schematic picture of the interhemispheric box model

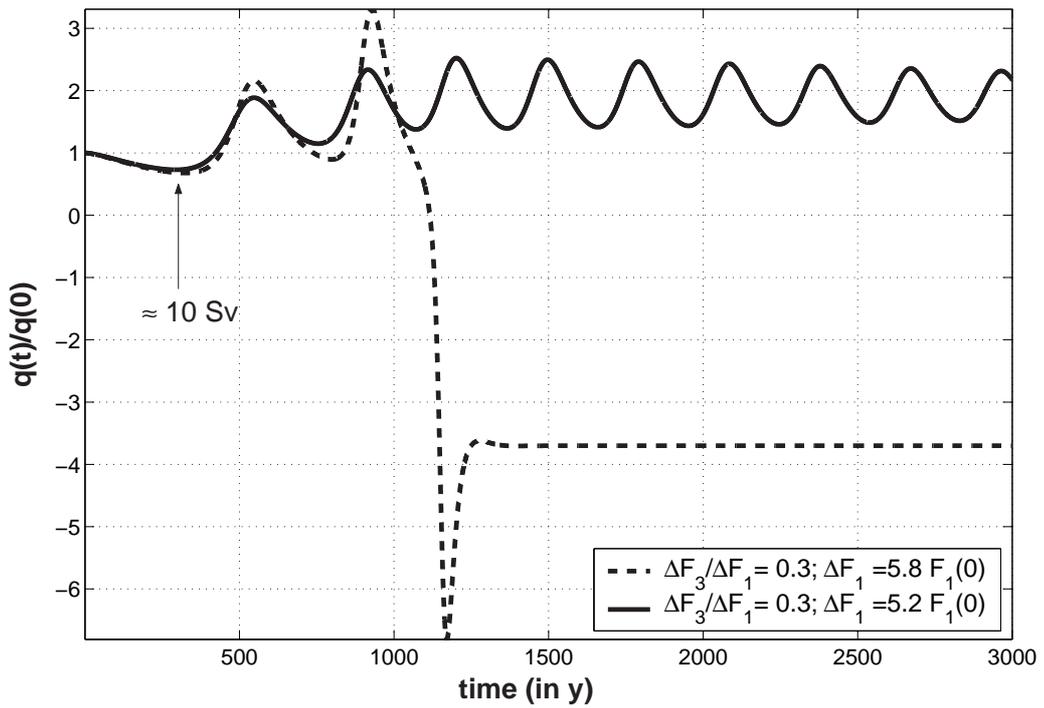

Figure 2: Evolution of the THC strength under a super- and sub-critical freshwater flux forcing

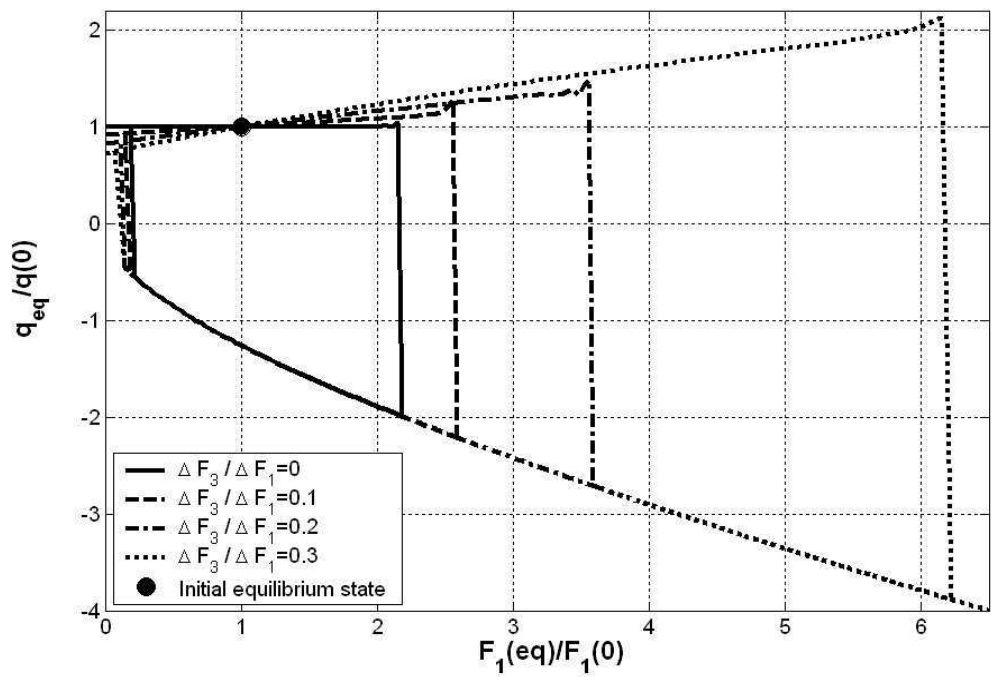

Figure 3: Hysteresis graph of freshwater flux forcings

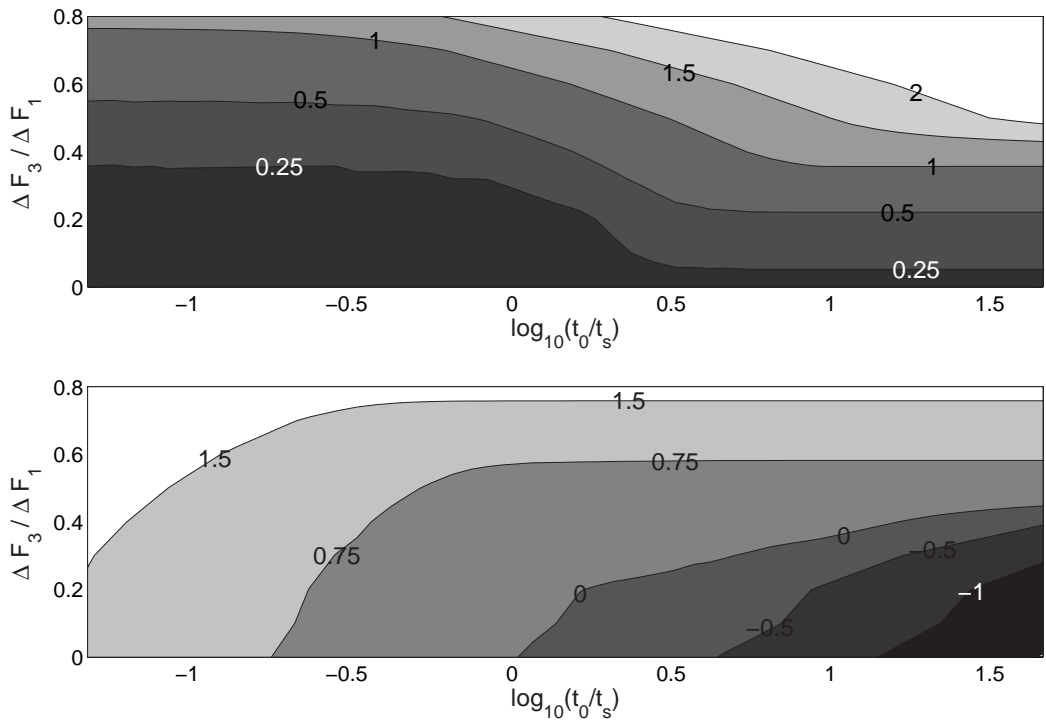

Figure 4: Stability of the THC against freshwater flux perturbations. a) Critical values of the total increase of the freshwater flux; contours of $\log_{10}[\Delta F_1/F_1(0)]$. b) Critical values of the rate of increase of the freshwater flux; contours of $\log_{10}[F_1^t \cdot t_0/F_1(0)]$ where $t_0$ is the total length of forcing

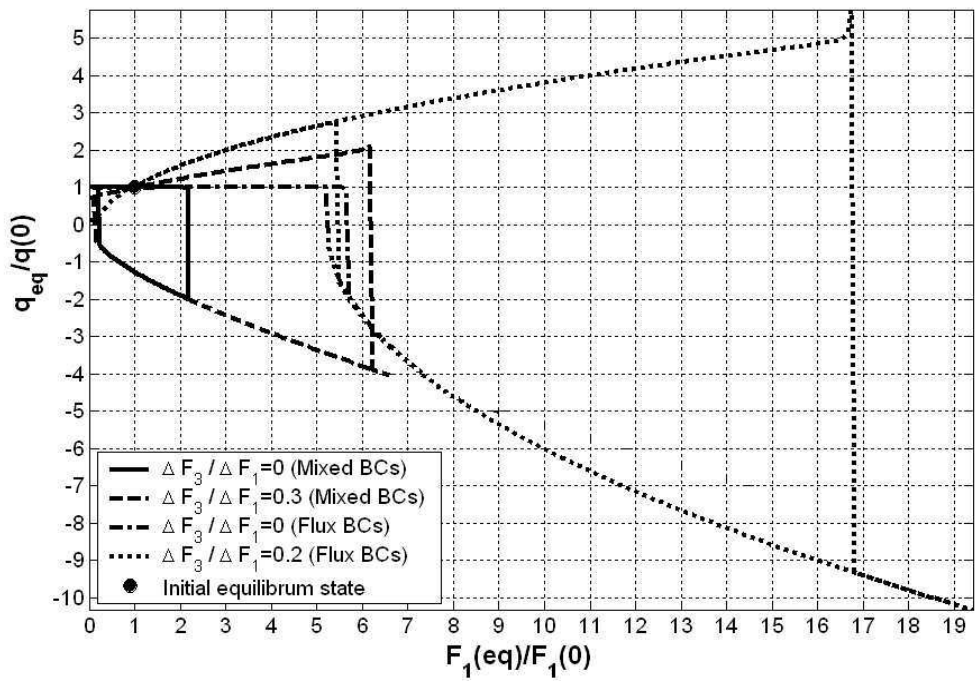

Figure 5: Comparison between the stability of the mixed and flux BCs uncoupled model to increases of the freshwater fluxes; hysteresis graphs

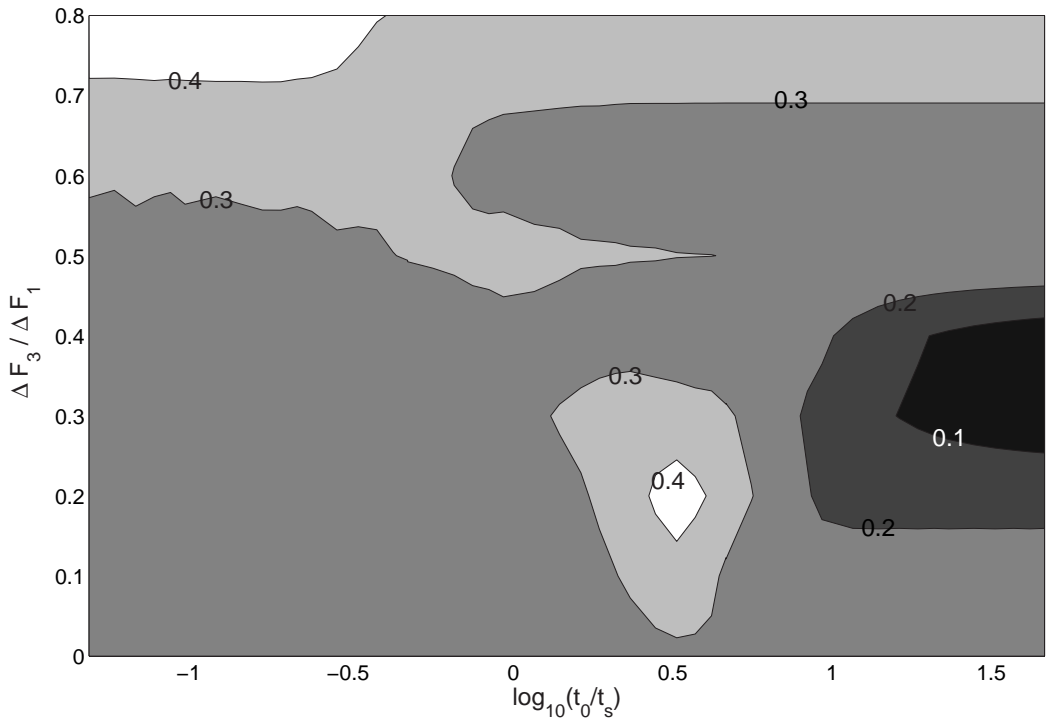

Figure 6: Comparison between the stability of the mixed and flux BCs uncoupled model to increases of the freshwater fluxes; contours of $\Delta F_1^{mix}/\Delta F_1^{flux}$

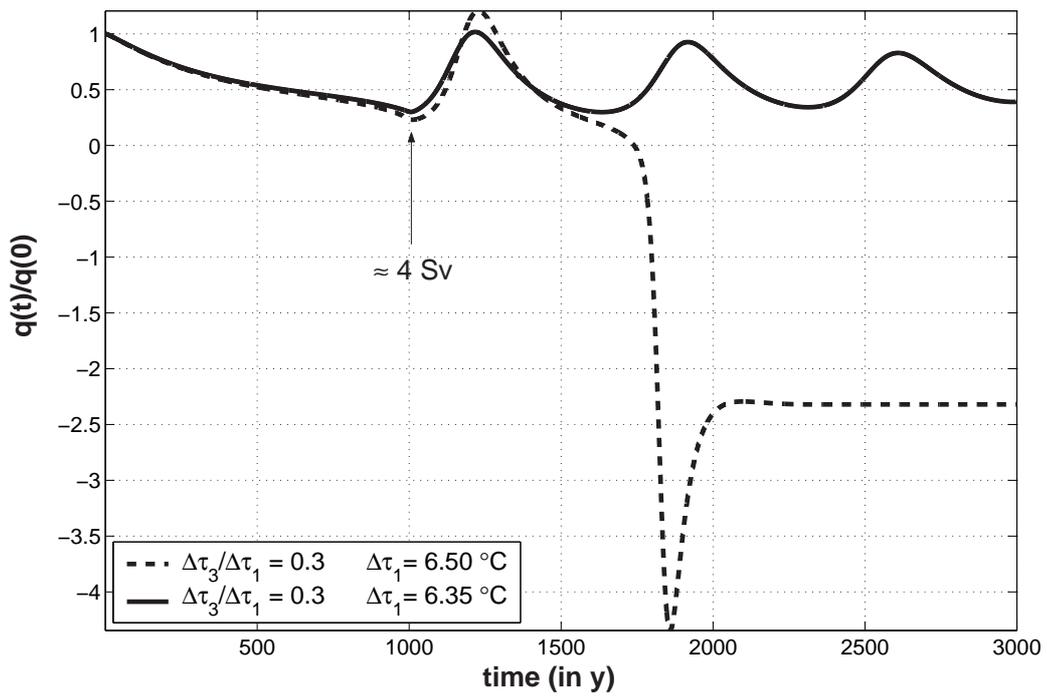

Figure 7: Evolution of the THC strength under a super- and sub-critical forcing to target temperatures

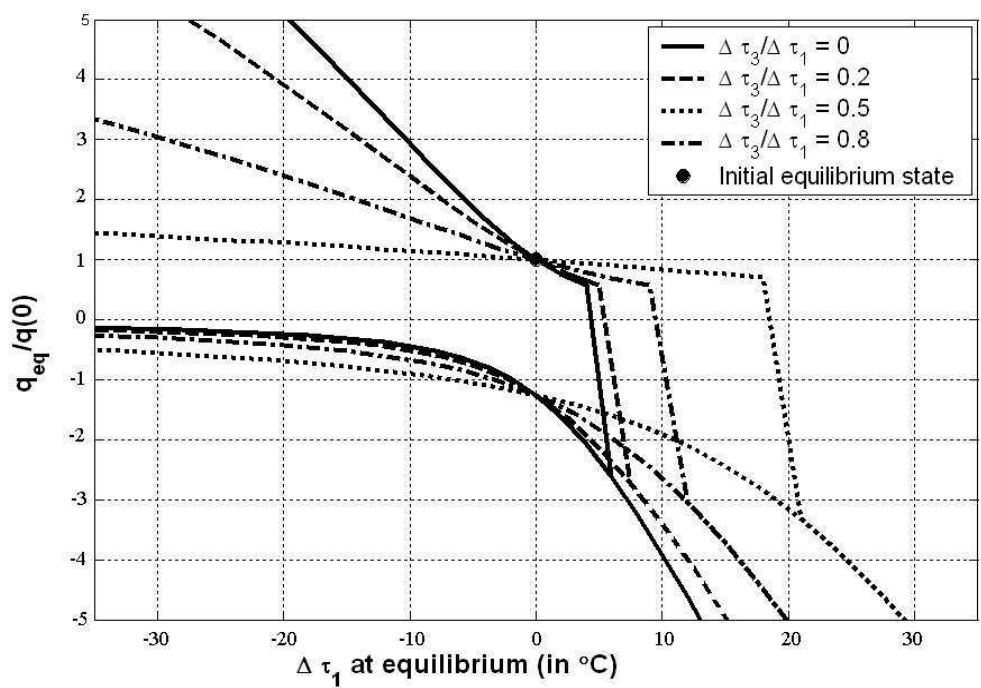

Figure 8: Hysteresis graph of the target temperature forcings

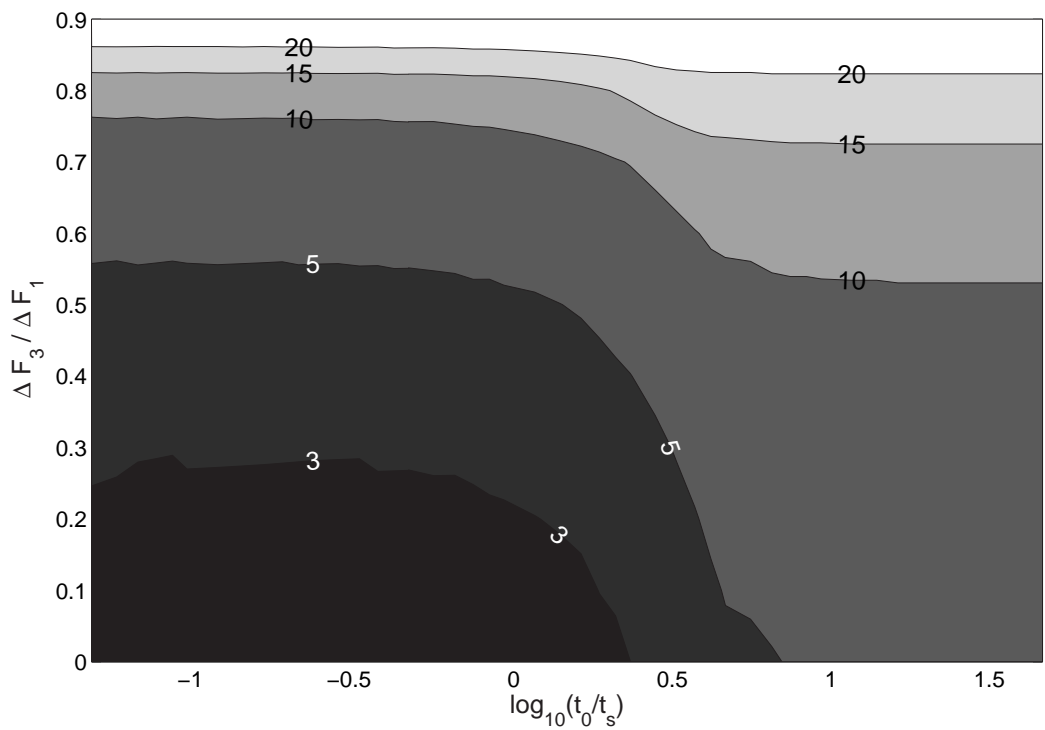

Figure 9: Stability of the THC against target temperature perturbations. Critical values of the total increase of the target temperature $\tau_1$; contours of $\Delta\tau_1$ in $°C$.